\documentclass[sigconf, nonacm]{acmart}

\usepackage{graphicx}
\graphicspath{{images/}}

\usepackage{booktabs}
\usepackage{array}
\usepackage{tabularx}
\usepackage{siunitx}
\usepackage{subcaption}
\newcolumntype{L}{>{\raggedright\arraybackslash}X}
\newcolumntype{C}{>{\centering\arraybackslash}X}

\usepackage{minted}
\newminted[pythoncode]{python}{
    fontsize=\small,
    frame=single,
    linenos=false,
    breaklines=true,
    breakanywhere=true
}
\newcommand{\todo}[1]{}
\renewcommand{\todo}[1]{{\color{red} TODO: {#1}}}

\newcommand{\step}[1]{\textbf{\texttt{step-#1}}}
\title{Scaling Hybrid Quantum--HPC Applications with the
  Quantum Framework}

\author{Srikar Chundury}
\email{schundu3@ncsu.edu}
\orcid{0009-0001-8335-9259}
\affiliation{%
  \institution{
    North Carolina State University
  }
  \city{Raleigh}
  \state{NC}
  \country{USA}
}

\author{Amir Shehata}
\email{shehataa@ornl.gov}
\orcid{0000-0002-2453-1426}
\affiliation{%
  \institution{
    Oak Ridge National Laboratory
  }
  \city{Oak Ridge}
  \state{TN}
  \country{USA}
}

\author{Seongmin Kim}
\email{kims@ornl.gov}
\orcid{0000-0001-5906-3004}
\affiliation{%
  \institution{
    Oak Ridge National Laboratory
  }
  \city{Oak Ridge}
  \state{TN}
  \country{USA}
}

\author{Muralikrishnan Gopalakrishnan Meena}
\email{gopalakrishm@ornl.gov}
\orcid{0000-0003-4048-4639}
\affiliation{%
	\institution{
    Oak Ridge National Laboratory
	}
	\city{Oak Ridge}
	\state{TN}
	\country{USA}
}

\author{Chao Lu}
\email{luc1@ornl.gov}
\orcid{0000-0001-7934-6933} 
\affiliation{%
    \institution{
        Oak Ridge National Laboratory
    }
    \city{Oak Ridge}
    \state{TN}
    \country{USA}
}

\author{Kalyana Gottiparthi}
\email{gottiparthik@ornl.gov}
\orcid{0000-0002-1354-0255}
\affiliation{%
	\institution{
    Oak Ridge National Laboratory
  }
	\city{Oak Ridge}
	\state{TN}
	\country{USA}
}

\author{Eduardo Antonio Coello Perez}
\email{coellopereea@ornl.gov}
\orcid{0000-0001-6074-0711}
\affiliation{%
	\institution{
    Oak Ridge National Laboratory
  }
	\city{Oak Ridge}
	\state{TN}
	\country{USA}
}

\author{Frank Mueller}
\email{fmuelle@ncsu.edu}
\orcid{0000-0002-0258-0294}
\affiliation{%
  \institution{
    North Carolina State University
  }
  \city{Raleigh}
  \state{NC}
  \country{USA}
}

\author{In-Saeng Suh}
\email{suhi@ornl.gov}
\orcid{0000-0002-6923-6455}
\affiliation{%
  \institution{
    Oak Ridge National Laboratory
  }
  \city{Oak Ridge}
  \state{TN}
  \country{USA}
}

\begin{document}

\begin{abstract}
Hybrid quantum-high performance computing (Q-HPC) workflows are
emerging as a key strategy for running quantum applications at scale
in current noisy intermediate-scale quantum (NISQ) devices. These
workflows must operate seamlessly across diverse simulators and
hardware backends since no single simulator offers the best
performance for every circuit type. Simulation efficiency depends
strongly on circuit structure, entanglement, and depth, making a
flexible and backend-agnostic execution model essential for fair
benchmarking, informed platform selection, and ultimately the
identification of quantum advantage opportunities.
In this work, we extend the Quantum Framework (QFw), a modular and
HPC-aware orchestration layer, to integrate multiple local backends
(Qiskit Aer, NWQ-Sim, QTensor, and TN-QVM) and a cloud-based quantum
backend (IonQ) under a unified interface. Using this integration, we
execute a number of non-variational
as well as variational workloads.
The results highlight workload-specific backend advantages: while Qiskit
Aer's matrix product state excels for large Ising models, NWQ-Sim not only
leads on large-scale entanglement and Hamiltonian but also shows
the benefits of concurrent subproblem execution in a distributed
manner for optimization problems. These findings demonstrate that
simulator-agnostic, HPC-aware orchestration is a practical path toward
scalable, reproducible, and portable Q-HPC ecosystems, thereby
accelerating progress toward demonstrating quantum advantage.

\end{abstract}

\keywords{Quantum computing, High-performance computing,
Quantum simulation, Hybrid quantum-classical workflows, Quantum frameworks}


\maketitle

\section{Introduction}
\label{sec:introduction}

As scientific and engineering problems increase in complexity,
traditional computational methods face fundamental limitations due to
exponential scaling~\cite{arute2019quantum, kim2022high}.  Domains
such as combinatorial optimization, materials discovery, and
large-scale linear algebra contain problem instances that become
classically intractable beyond modest
sizes~\cite{montanaro2016quantum, kim2024wide, lu2025lugo}.  Quantum
computing offers a promising paradigm to address such challenges by
exploiting quantum parallelism, entanglement, and
superposition~\cite{google2025quantum}.

Progress in quantum hardware~\cite{koch2007charge, arute2019quantum,
  monroe2013scaling, freedman2003topological} has delivered small- to
mid-scale quantum devices, so-called quantum processing units (QPUs),
that can act as accelerators alongside CPUs and GPUs in
high-performance computing (HPC) environments.  Yet practical quantum
advantage remains out of reach due to limited qubit counts, device
noise, and the need for seamless integration with large-scale
classical workflows~\cite{preskill2018nisq}.  These constraints have
intensified interest in hybrid quantum-classical algorithms, where
quantum subroutines are embedded within classical outer loops.  In
contrast to their non-variational counterpart, variational algorithms
are less prone to adverse affects of today's noisy quantum
devices. Such algorithms, including the Quantum Approximate
Optimization Algorithm
(QAOA)~\cite{farhi2014quantumapproximateoptimizationalgorithm,
  Farhi_2022}, the Variational Quantum Linear Solver
(VQLS)~\cite{Bravo_Prieto_2023}, and the Variational Quantum Eigen
solver \cite{peruzzo2014variational}, leverage the quantum-classical
hybrid structure, coupling near-term QPUs with HPC-scale classical
optimization.

To evaluate the capabilities of emerging Q-HPC systems, both
non-variational and variational workloads should be considered.  The
non-variational set that we choose to analyze includes the generation
of Greenberger–Horne-Zeilinger
(GHZ)~\cite{greenberger2007goingbellstheorem} fully entangled states,
Hamiltonian simulation (HAM)~\cite{Ising1925}, the transverse-field
Ising model (TFIM)~\cite{Fradkin2013}, and the Harrow–Hassidim-Lloyd
(HHL)~\cite{Harrow_2009} linear solver.  The variational set includes
QAOA~\cite{kim2024dqaoa}, and Distributed QAOA
(DQAOA)~\cite{kim2025distributed}.  While these workloads represent
key algorithmic families in physics, engineering, and scientific
computing~\cite{Low_2019, Uvarov_2020, Peruzzo_2014,
  anschuetz2018variational}, they are computationally demanding, and
hence are seldom evaluated at large scales.

To address the scalability issue, we extend the Quantum Framework
(QFw)~\cite{osti_2498439,
  xu2025gpuaccelerateddistributedqaoalargescale,
  shehata2024frameworkintegratingquantumsimulation}, a portable and
HPC-aware orchestration layer for quantum applications, to integrate
multiple local and one cloud-based backend under a unified interface.
The local backends include NWQ-Sim~\cite{li2021svsim}, Qiskit
Aer~\cite{Qiskit}, TN-QVM~\cite{mccaskey2016tnqvm}, and
QTensor~\cite{lykov2021qtensor}, each launched in distributed mode via
the PMIx Reference RunTime Environment (PRTE) and the Message Passing
Interface (MPI). The cloud backend interfaces with
IonQ~\cite{ionq_cloud_jobs}.  Deployed on the Frontier
supercomputer~\cite{frontier_cluster} via SLURM heterogeneous job
groups, QFw enables coordinated execution across heterogeneous
resources with fine-grained control over scheduling, communication,
and device allocation.

Furthermore, we show that QFw can run identical application code
across all backends without modification, supporting reproducible
benchmarking and rapid backend substitution. Our evaluation reveals
workload-specific performance advantages and demonstrates the role of
distributed execution and asynchronous orchestration in scaling large
variational workloads.

The contributions of this work are the following:
\begin{itemize}
    \item We integrate Qiskit-Aer, QTensor, and IonQ backends into QFw.
    \item We scale execution and evaluation of variational and
      non-variational workloads on QFw across multiple local
      simulators and a cloud backend.
    \item We provide an implementation together with large-scale
      experiments of DQAOA for metamaterial optimization on the
      Frontier supercomputer.
\end{itemize}

The remainder of this paper is organized as follows:
Section~\ref{sec:background} reviews technical background and related work;
Section~\ref{sec:methodology} details backend integration and orchestration design;
Section~\ref{sec:experimental_setup} describes the compute platform, backends, and benchmarks;
Section~\ref{sec:results} presents performance and scalability results;
Section~\ref{sec:discussion} analyzes observations and implications;
and Section~\ref{sec:conclusion} summarizes our findings and outlines
future directions.

\section{Background}
\label{sec:background}

This section provides the context for our
study. We first summarize the QFw execution model and services that
enable distributed, backend-agnostic orchestration
(Section~\ref{sec:qfw}).  We then define the workload taxonomy,
covering fixed-circuit non-variational kernels (GHZ, HAM/TFIM, and
HHL, see Section~\ref{sec:background_non_variational_workloads}) and
parameterized, optimizer-in-the-loop variational algorithms (QAOA and
DQAOA, see Section~\ref{sec:background_variational_workloads}).  Together, these elements motivate the
orchestration choices used in our experiments and ground the
performance analysis presented later.

\subsection{Quantum Framework (QFw)}
\label{sec:qfw}

The QFw~\cite{osti_2498439,
  xu2025gpuaccelerateddistributedqaoalargescale,
  shehata2024frameworkintegratingquantumsimulation} is a scalable
orchestration platform designed to integrate quantum and classical
computing resources for large-scale hybrid execution. QFw provides a
modular, backend-agnostic interface for managing quantum circuit
execution across CPUs, GPUs, and QPUs in HPC environments. By
abstracting backend-specific details, QFw enables quantum applications
to remain portable and reproducible, allowing researchers to switch
between simulation and hardware targets with minimal code changes.

QFw builds on the Process Management Interface for Exascale (PMIx)
Reference RunTime Environment (PRTE) in the Distributed Virtual
Machine (DVM) mode to enable rapid process spawning and low-latency
coordination across distributed nodes. Its architecture comprises
three primary services. (1) The \emph{Quantum Platform Manager} (QPM)
acts as a central dispatcher, selecting execution backends and
managing task configurations. (2) The \emph{Quantum Resource
  Controller} (QRC) schedules and launches quantum tasks across MPI
ranks, ensuring efficient utilization of allocated resources. (3)
Communication between these components is handled by the
\emph{Distributed Execution Framework} (DEFw), a lightweight remote
procedure call (RPC) layer optimized for HPC-scale deployments.

Through this design, QFw supports a variety of quantum backends,
including state-vector and tensor-network simulators. Users typically
access QFw through a Python API module (\texttt{QFwBackend}) that
provides a drop-in backend compatible with frameworks such as
Qiskit~\cite{javadi2024quantum} and
PennyLane~\cite{bergholm2018pennylane}.  This interface translates
circuit execution requests into QFw API calls, which are then
dispatched to the selected backend. Simulator workloads are executed
in distributed mode via MPI, while hardware requests are routed to the
appropriate service interface. In both cases, results are returned in
standardized formats, shielding applications from backend-specific
parsing.

By combining HPC process management, distributed task scheduling, and
backend-agnostic execution, QFw serves as a bridge between quantum and
classical ecosystems~\cite{shehata2025bridginghpcquantum}. It provides
a unified foundation on which scalable, reproducible, and portable
quantum workflows can be developed and deployed on leadership-class
computing platforms.

\subsection{Non-Variational Workloads}
\label{sec:background_non_variational_workloads}

Non-variational workloads execute fixed circuits without a classical
parameter-update loop and therefore expose the raw execution, memory,
and communication characteristics of each backend. We consider (i) GHZ
state preparation, which stresses long-range entanglement growth with
shallow but highly correlated
circuits~\cite{greenberger2007goingbellstheorem}; (ii) Hamiltonian
simulation of Ising-type models, including TFIM, which exercises
trotterized time evolution and controlled two-qubit interactions with
tunable depth~\cite{Ising1925,Fradkin2013,Low_2019}; and (iii) the HHL
linear solver, which illustrates deeper coherent subroutines (e.g.,
phase estimation and controlled rotations) and ancilla
management~\cite{Harrow_2009}. These kernels span distinct structures
and depths, making them informative for contrasting state-vector and
tensor-network simulators, including their distributed
implementations~\cite{li2021svsim,lykov2021qtensor,mccaskey2016tnqvm}.
Within QFw, identical circuits are issued through standard frontends
and dispatched uniformly to multiple local simulators and the cloud
backend, enabling portable and reproducible
comparisons~\cite{javadi2024quantum,bergholm2018pennylane}.

\subsection{Variational Workloads}
\label{sec:background_variational_workloads}

Variational workloads couple parameterized quantum circuits with a
classical optimizer, repeatedly preparing, measuring, and updating
parameters to minimize a task-specific objective~\cite{Peruzzo_2014}.
Our focus is on QAOA, whose layered cost–mixer ansatz provides a
controllable depth-quality trade-off and well-studied scaling behavior
on combinatorial problems~\cite{Farhi_2022}. In particular, our work
focuses on finding novel meta-materials by formulating the optimization
problem as a quadratic unconstrained binary optimization (QUBO).

\subsubsection*{Distributed Quantum Approximate Optimization Algorithm (DQAOA)}
\label{sec:background_dqaoa}

The DQAOA is an extension of QAOA designed to address the hardware and
scale limitations of near-term quantum computing systems. In this
approach, a large combinatorial optimization problem is decomposed
into smaller sub-problems, each requiring much fewer qubits and
shallower circuits. These sub-problems can be executed simultaneously
on multiple quantum simulators or hardware, where their results are
combined to form an approximate global solution. Such decomposition
mitigates two key constraints in current quantum systems: (1) limited
qubit counts and (2) circuit depth restrictions due to noise.

DQAOA naturally lends itself to HPC integration, as many sub-problems
can be solved in parallel. A typical workflow allocates classical and
quantum resources, dispatches sub-problem circuits concurrently, and
aggregates their outputs for classical post-processing. This parallel
execution is well-suited for batch schedulers and distributed
execution frameworks, enabling substantial reductions in wall time for
large-scale problems.

Kim~\emph{et~al.}~\cite{kim2024dqaoa} proposed DQAOA that coupled HPC
resources with quantum backends, distributing sub-problem execution
across heterogeneous compute nodes. Their study showed that this
approach can handle problems with up to thousands of binary variables
with high solution quality.
Xu~\emph{et~al.}~\cite{xu2025gpuaccelerateddistributedqaoalargescale}
extended this model to a GPU-accelerated implementation using Qiskit
Aer and MPI-based orchestration, achieving significant performance
gains on leadership-class supercomputers.

These studies illustrate how DQAOA can combine quantum parallelism
with classical HPC scalability. By exploiting distributed resources at
both the algorithmic and simulation levels, DQAOA provides a pathway
toward solving larger optimization instances than would be possible on
quantum computing alone.

In this context, iterative workloads amplify orchestration demands
(many circuit evaluations per optimization step) and therefore benefit
directly from QFw's backend-agnostic scheduling and distributed
execution model.  Using a unified frontend (e.g., Qiskit or PennyLane)
with QFw allows the same QAOA/DQAOA application code to target
multiple simulators and a cloud backend without modification,
supporting fair benchmarking and rapid platform
substitution~\cite{javadi2024quantum,bergholm2018pennylane}.

\section{Related Work}
\label{sec:related_work}

Early efforts to integrate quantum acceleration into classical HPC
workflows have focused on bridging existing cluster tools with
emerging quantum resources. For instance, Esposito and
Haus~\cite{esposito2025slurmheterogeneousjobshybrid} leverage SLURM's
heterogeneous jobs feature and MPI to co-schedule hybrid
classical-quantum workflows on supercomputers, aiming to eliminate
idle time on scarce quantum processors.  Their method requires
refactoring a monolithic hybrid program into a sequence of smaller
jobs, each encapsulating one quantum computation alongside its
surrounding classical processing. This allows the quantum device to be
released after each quantum step and reused by other jobs,
significantly improving utilization. While effective, this approach
demands moderate user intervention (splitting code and managing data
hand-offs) and relies on the HPC scheduler to interleave tasks,
without deeper automation or resource awareness beyond job queuing.

More automated scheduling frameworks have been proposed to manage
quantum tasks in HPC environments. SCIM MILQ by Seitz \emph{et
  al.}~\cite{Seitz_2024} is an HPC quantum scheduler that combines
conventional job scheduling techniques with quantum-specific methods
like circuit cutting. By partitioning large quantum circuits and
intelligently scheduling the pieces, SCIM MILQ seeks to minimize
overall makespan and noise impact when running multiple quantum tasks
on limited hardware. This is a step toward specialized runtime systems
that can optimize hybrid workloads, although such research prototypes
have yet to be integrated into general-purpose HPC job
managers. Similarly,
XACC~\cite{mccaskey2019xaccsystemlevelsoftwareinfrastructure} provides
a low-level software infrastructure for heterogeneous
quantum-classical computing, exposing a service-oriented framework for
quantum program compilation and execution across diverse hardware
backends. XACC's hardware-agnostic interfaces lay the groundwork for
tightly coupling quantum co-processors with classical code, but
require adoption within HPC toolchains.

Another line of work emphasizes unified software stacks to hide
quantum-classical complexity from end users. The Munich Quantum
Software Stack (MQSS)~\cite{qdessi_munich_quantum_valley} being
developed by TUM and LRZ is a comprehensive platform that deeply
integrates quantum devices into HPC centers. MQSS provides a single
access point to multiple quantum backends at the LRZ supercomputing
facility, accessible via a web portal, command line, or through hybrid
jobs tied into the cluster scheduler~\cite{qexa_project}. The stack
automates the workflow from user-submitted quantum programs to
execution on available hardware, including dynamic compilation that
adapts to device calibration data~\cite{mqt}. Our work shares a
similar goal of enabling flexible backend selection and seamless
quantum integration, but MQSS is tailored to a specific institutional
environment.

Beyond HPC-centric solutions, the rise of cloud quantum services also
provides relevant context. Platforms such as Amazon
Braket~\cite{braket}, Microsoft Azure
Quantum~\cite{microsoft_azure_quantum}, and qBraid~\cite{qbraid}
aggregate a variety of quantum processors (superconducting, ion-trap,
etc.) from multiple vendors and offer unified interfaces for job
submission. These services let users develop algorithms with familiar
open-source frameworks (e.g., Qiskit, Cirq, PennyLane) and run them on
different backends through a common API. For example, qBraid's
environment enables one-click access to over two dozen quantum devices
across providers, with cross-framework compatibility and centralized
job tracking. Such cloud platforms demonstrate the value of backend
flexibility and ease of use, though they operate in remote data
centers rather than tightly coupled to local HPC resources.
Conqure~\cite{mahesh25-2} combines an Amazon-compatible interface with
a Qiskit front-end and Slurm-based scheduling coordinated with HPC in
an open-source environment aimed at lab-hosted quantum devices and
local or remote simulation with a private cloud presence.

In our work, we similarly support multiple quantum
backends---combining both local HPC ones and a remote one
under a single workflow. However, we differ by focusing on hybrid
workflow performance and co-scheduling in an HPC context, rather than
pure cloud access. All prior art has either required users to manually
split and schedule quantum tasks or has been limited to specific
ecosystems, with only small-scale simulation. Our approach strives to
generalize hybrid quantum-classical scheduling with minimal user
effort. It provides a unified framework that dynamically selects
quantum backends and orchestrates their interaction with classical
computations, thereby advancing the state of the art in flexible
quantum workflow management.

\section{Methodology}
\label{sec:methodology}

QFw addresses two complementary goals: (i) providing a clean, modular
path to integrate new simulators and hardware backends, and (ii)
enabling an application execution model that scales across HPC
resources and hybrid cloud endpoints with minimal changes to user
code.

QFw exposes a \emph{frontend API} to applications and a \emph{backend
  QPM API} to simulators or hardware.  QPM coordinates a pool of
worker processes, each bound to an HPC core, and dispatches batches of
quantum circuit tasks either via PRTE/MPI for on-premise execution or
through REST for cloud endpoints. Results are returned asynchronously
to the frontend, ensuring non-blocking execution for iterative and
variational workflows.

\begin{figure*}[t]
  \centering
  \fbox{\includegraphics[width=\textwidth]{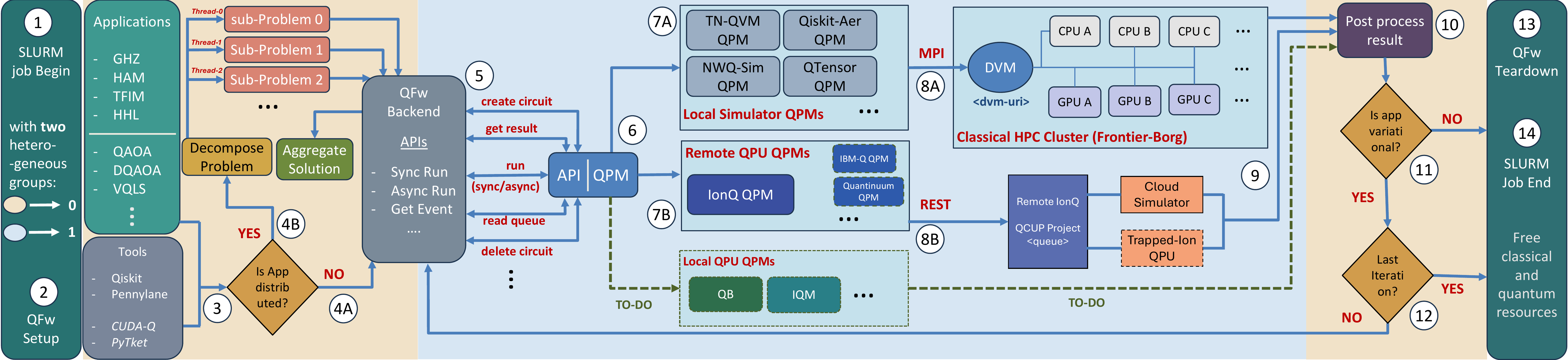}}
  \caption{QFw architecture and execution flow. QPM manages worker
  pools, dispatches batches to local MPI ranks or cloud endpoints, and
  streams results back to applications.~\textit{Note: Background shading
  corresponds to the heterogeneous groups as pointed out in \step{1},
  with the warmer tone
  indicating hetgroup-0 components and the cooler tone indicating
  hetgroup-1 components.}}
  \label{fig:qfw_workflow}
\end{figure*}

The end-to-end workflow is illustrated in Fig.~\ref{fig:qfw_workflow}.
In \step{1}, a SLURM job is submitted with two heterogeneous groups:
one for the application (hetgroup-0) and another for QFw-related management and
simulator nodes (hetgroup-1).
In \step{2}, the QFw setup procedure launches the core infrastructure,
including the DEFw with a PRTE/DVM URI shared across all processes,
multiple QPM processes for managing quantum circuit execution jobs and
queues, and multiple QRC worker threads
per QPM process for triggering MPI runs on the nodes allocated to QFw
(hetgroup-1). Only the first node in this group houses
QFw's management services, while all nodes, including the first, act
as workers. In \step{3}, the user application, implemented in
frameworks such as Qiskit or PennyLane, is executed.

If the
application is distributed (\step{4A}), the programmer may employ MPI
(e.g., via \texttt{mpi4py}) or use multithreading within the
application's node group and
must connect to the same backend API specified in \step{5}. If it is
not distributed (\step{4B}), the QFwBackend is invoked directly,
connecting to the QPM API in hetgroup-1 and preparing to submit
circuits
to any QPM accessible to the backend. In \step{5}, the QFwBackend,
implemented as a Qiskit BackendV2 interface, routes application
requests to the designated QPM service via RPC calls handled by
DEFw. These calls, which include circuit creation, deletion,
execution, and status queries, are processed in \step{6} by the
appropriate QPM implementation depending on the application's
configuration. Circuit execution (\step{7}) may occur on local HPC
resources in hetgroup-1 or on remote QPUs via REST calls.

In \step{9},
execution results are marshaled into the common QPM API format before
being forwarded for application-level post-processing
(\step{10}). Depending on the application logic (\step{11}), e.g.,
whether additional iterations for a classical optimizer are required,
control is either returned to the QFwBackend for continued execution
or the application concludes (\step{12}). In \step{13}, QFw performs a
controlled teardown in hetgroup-1, shutting down RPC services,
releasing
worker allocations, and freeing any remote QPU queue reservations
(currently using QCUP's shared queue). Finally, the SLURM job
terminates in \step{14}.

The next two subsections detail (1) how new simulators or hardware
backends plug into QFw, and (2) how applications interface with the
frontend API to submit batched circuits and retrieve asynchronous
results.

\subsection{Integrating Backends}
\label{sec:integrating_backends}

Integrating a new backend into QFw requires implementing a
\emph{Backend QPM} that conforms to a predefined
QPM-API. Consequently, each backend that is to be integrated with QFw
as a new QPM should (1) accept a standardized circuit/problem
description, (2) configure backend-specific runtime parameters, (3)
launch execution via MPI/PRTE (on-prem) or REST (cloud), and (4)
marshal results into QFw's unified return format. Hence, the frontend
and backend are cleanly decoupled and the application code remains
unchanged when swapping backends or retargeting sub-backends. QFw also
unifies logging, error propagation, and timing instrumentation so QPM
can maintain comparable per-backend performance profiles.

QFw currently supports five backends spanning state-vector and
tensor-network simulation, plus a cloud QPU provider. Each integration
reflects challenges in terms of robustness, scaling model, and ease of
deployment on HPC systems:

\begin{itemize}
\item \textbf{TN-QVM (ORNL).} TN-QVM is a tensor-network simulator
  that uses the ExaTN library underneath.
  TN-QVM can target multiple topologies (MPS, TTN, PEPS). We wrote a
  thin wrapper to select the topology as a sub-backend. In QFw, we
  currently support and test \emph{ExaTN-MPS}.

\item \textbf{NWQ-Sim (PNNL).} The state-vector engine (SV-Sim) is fully
  integrated.  Historically, SV-Sim offers several sub-backends
  (OpenMP, MPI, CPU, AMD GPU) that are supported and selectable at
  runtime. NWQ-Sim's native MPI distribution makes it a good fit for
  multi-node CPU/GPU HPC runs.

\item \textbf{QTensor (ANL).} QTensor is a tree tensor-network
  approach built around the \texttt{qtree} library. Although QTensor
  is primarily designed for QAOA-style workloads (emphasizing
  expectation estimation on sparse QUBOs via light-cone style
  optimizations), QFw uses it for full-state contraction through
  \texttt{qtree}. QFw has been
  tested thoroughly with \texttt{numpy}. MPI distribution is enabled
  via \texttt{mpi4py}.

\item \textbf{Qiskit Aer.} Aer remains a state-of-the-art single-node
  circuit simulator and provides several sub-backends:
  \texttt{statevector}, \texttt{matrix\_product\_state},
  \texttt{stabilizer}, and \texttt{automatic} (which selects among
  others). MPI support exists via chunking across ranks.  QFw has been
  exercised with \emph{mps}, \emph{statevector}, and
  \emph{automatic}. GPU acceleration is available (CUDA by default);
  HIP/ROCm requires a custom Aer build for AMD GPUs.

\item \textbf{IonQ (cloud).} For the cloud path, simple REST suffices.
  In practice, we leverage IonQ's Qiskit \texttt{BackendV2} plugin,
  which handles the REST plumbing but adds helpful
  boilerplate. This work extensively tested the \emph{simulator} sub-backend.
\end{itemize}

Table~\ref{tab:simulators} summarizes capabilities and caveats. In all
cases, backends and sub-backends are selected via lightweight runtime
properties (e.g., \texttt{\{``backend'': ``qtensor'', ``subbackend'':
  ``numpy''\}}), so users can shift between engines without
refactoring their algorithms. The key contribution here is the ease of
integrating new backends and combing components to this end.
What's more, the same QFw backend object can be plugged into popular
frameworks (e.g., Qiskit, PennyLane); swapping
\texttt{backend}/\texttt{subbackend} toggles engines and device
targets without changing the user's quantum program as illustrated in
Figure~\ref{lst:qfw_code}.
\noindent
\begin{figure}[ht]
\begin{pythoncode}
# Qiskit + QFw example
from qfw_qiskit import QFwBackend
backend = QFwBackend(
    properties = {
        "backend":"nwqsim",
        "subbackend":"MPI"
    }
)
# Pennylane + QFw example
import pennylane as qml
dev = qml.device('qiskit.remote', wires=2,
     backend=backend)
\end{pythoncode}
\caption{Selecting the QFwBackend.}
\label{lst:qfw_code}
\end{figure}

\newcolumntype{P}[1]{>{\raggedright\arraybackslash}p{#1}} 
\newcolumntype{C}[1]{>{\centering\arraybackslash}p{#1}} 
\begin{table*}[t]
  \centering
  \caption{Backends used with QFw.}
  \label{tab:simulators}
  \begin{tabularx}{\linewidth}{C{0.10\linewidth} P{0.15\linewidth} C{0.04\linewidth} C{0.04\linewidth} C{0.09\linewidth} L}
    \toprule
    \textbf{Backend} & \textbf{Sub-backend(s)} & \textbf{CPU}
    & \textbf{GPU} & \textbf{Native MPI} & \textbf{Notes} \\
    \midrule
    TN\mbox{-}QVM (ORNL) &
      \begin{tabular}[t]{@{}l@{}}
        exatn-mps\\
        TTN (pending)\\
        PEPS (planned)
      \end{tabular} &
      Yes & Yes\textsuperscript{*} & Yes\textsuperscript{*} &
      Tensor-network simulator; QFw wrapper selects topology. QFw is tested with \emph{ExaTN-MPS}. TTN currently blocked by \texttt{.xasm} vs.\ \texttt{.qasm}; PEPS is architecturally supported. \\
      \midrule
      NWQ\mbox{-}Sim (SV\mbox{-}Sim) &
      \begin{tabular}[t]{@{}l@{}}
        OpenMP\\
        MPI\\
        CPU\\
        AMDGPU
      \end{tabular} &
      Yes & Yes & Yes &
      Fully integrated. At the time of development, HIP+MPI lacked complete upstream support; other sub-backends are supported via QFw. \\
      \midrule
      Qiskit Aer &
      \begin{tabular}[t]{@{}l@{}}
        statevector\\
        matrix\_product\_state\\
        automatic
      \end{tabular} &
      Yes & Yes\textsuperscript{\dag} & Yes &
      Strong single-node performance; MPI uses chunking. QFw tested with \emph{mps}, \emph{statevector}, and \emph{automatic}. HIP/ROCm requires a custom Aer build. \\
      \midrule
      QTensor (ANL) &
      \begin{tabular}[t]{@{}l@{}}
        numpy\\
        cupy (planned)\\
        pytorch (planned)
      \end{tabular} &
      Yes & Yes & Yes &
      Tree TN (\texttt{qtree}). Designed for QAOA expectation estimation on sparse QUBOs, but used in QFw for full-state contraction; currently tested thoroughly with \texttt{numpy}. MPI via \texttt{mpi4py}. \\
      \midrule
      IonQ (cloud) &
      \begin{tabular}[t]{@{}l@{}}
        simulator\\
        hardware (planned)
      \end{tabular} &
      \multicolumn{3}{c}{N/A} &
      Integrated via IonQ's Qiskit \texttt{BackendV2} plugin (REST under the hood). \\
      \bottomrule
  \end{tabularx}
  {\footnotesize \textsuperscript{*}\,Engine-dependent via ExaTN build options.\quad
  \textsuperscript{\dag}\,CUDA by default; HIP/ROCm requires a custom Aer build.}
\end{table*}

\subsection{Integrating Applications}
\label{sec:integrating_apps}

QFw is deployed on HPC systems via SLURM's heterogeneous job
allocation model.  A typical launch reserves two resource groups:
\texttt{hetgroup-0} for the application's classical control logic,
and \texttt{hetgroup-1} for QFw services and backend execution. The QPM
service is started on the lead node of \texttt{hetgroup-1},
instantiating a PRTE-DVM that spans all backend nodes. This separation
ensures that backend execution is isolated from application
orchestration overheads and allows for independent scaling of the
classical and quantum-simulation components.

On the frontend, applications interact with QFw through a
lightweight
Python client, \texttt{QFWBackend}, which exposes a uniform
\texttt{execute()} method. Applications generate circuits or problem
instances using their preferred SDK (e.g., Qiskit) and pass them
directly to QFw. Execution calls are translated into RPCs to QPM,
which schedules them across worker pools.

For \textbf{non-variational workloads} (GHZ, HAM, TFIM, HHL), QFw
batches independent circuit instances across available cores,
maximizing throughput. For \textbf{variational workloads} (QAOA,
DQAOA), the application issues multiple asynchronous calls within each
optimization iteration, enabling parameter sweeps without idle time.


To enable scalable quantum optimization workflows, we extended QFw to
\textbf{DQAOA applications}.  The workflow
(Fig.~\ref{fig:qfw_workflow}) begins with a dual-group heterogeneous
SLURM allocation: one group dedicated to the DQAOA application and
another to QFw services. In the QFw group, a unique DVM-URI is created
and QPM services are launched. On the DQAOA side, a Qiskit-based
Python application initializes with a QUBO matrix as input. During
initialization, the QFwBackend establishes a secure connection to the
QPM services.

The large QUBO is decomposed into multiple subQUBOs using either random
partitioning or decomposition methods  directed by an impact factor.
The application uses Python's \texttt{threading} module to issue
concurrent \texttt{solve} calls for each subQUBO, as the workload is
primarily I/O-bound. For local HPC simulators, these calls trigger
asynchronous RPCs to QFw, which dispatch the jobs via PRTE and launch
MPI-parallel circuit executions across the allocated nodes. For remote
quantum services such as the IonQ cloud simulator, QPM translates the
request into REST API calls, managing job submission and result
retrieval transparently.

As subQUBO results return, they are aggregated to update the
classical optimizer, and the cycle repeats until convergence. Upon
completion, QPM services are terminated, the DVM-URI is released, and
SLURM resources are deallocated. This integration design unifies local
MPI-based simulation and remote cloud backends under a single
asynchronous workflow, enabling DQAOA to scale efficiently across
heterogeneous resources while preserving full backend flexibility.



This modular integration model enables new backends and applications
to be combined arbitrarily, supporting reproducible, large-scale
quantum workloads without rewriting core application logic.

\section{Experimental Setup}
\label{sec:experimental_setup}

\paragraph{Compute Platform:}
The experiments are run on a Frontier test cluster at OLCF with 32
nodes. Each Frontier compute node contains one 64-core ``Optimized 3rd
Gen'' AMD EPYC CPU with \SI{512}{\gibi\byte} DDR4 memory, and four AMD
Instinct MI250X GPUs. Each MI250X comprises two GCDs (8 logical GPUs
per node), each with \SI{64}{\gibi\byte} HBM2e. Nodes are connected
via HPE Slingshot~200 with an aggregate node-injection bandwidth of
\SI{800}{\giga\bit\per\second}. Each node exposes eight last-level
cache (LLC) domains, each serving eight CPU cores; to minimize OS
noise, we reserve one core per LLC for kernel/system processes,
leaving \textbf{56 application cores per node} available to QFw.

\paragraph{Job Orchestration, Builds, and Protocol:}
We use SLURM heterogeneous job groups: \emph{hetgroup-0} hosts the
application layer and \emph{hetgroup-1} hosts Quantum Framework (QFw)
simulation workers. The QPM (on the first node of \emph{hetgroup-1})
spawns eight worker threads and distributes circuit-execution tasks
round-robin, using PRTE across \emph{hetgroup-1} for local simulators or
REST API calls for IonQ cloud backends. All simulators/backends are built
from their latest public releases
with \texttt{-O3}; GPU builds are enabled when available (e.g.,
cuQuantum/GPU options for Aer).  Each experiment is repeated three
times (limited by allocation) for which we report the mean and
standard deviation. Unless stated otherwise, one node is reserved for
the application layer (\emph{hetgroup-0}) while QFw scales on
\emph{hetgroup-1}.

\paragraph{Backends:}
We use the set of simulators and the IonQ cloud emulator (with
hardware executions planned for future work) described in
Section~\ref{sec:integrating_backends} and Table~\ref{tab:simulators}.

\paragraph{Benchmarks:}
We evaluate both non-variational and variational applications as
described in Section~\ref{sec:integrating_apps}; problem sizes are
enumerated in Table~\ref{tab:benchmarks}.

\begin{table}[t]
\centering
\caption{Benchmarks and problem sizes grouped by category. QAOA
  reports \emph{QUBO size}. DQAOA reports \emph{QUBO size} and subQUBO
  settings as \emph{(subqsize\textsuperscript{*},
    nsubq\textsuperscript{\dag})}.}
\label{tab:benchmarks}
\setlength{\tabcolsep}{3pt}
\small
\begin{tabular}{@{}%
  >{\raggedright\arraybackslash}p{.40\columnwidth}%
  >{\raggedright\arraybackslash}p{.58\columnwidth}@{}}
\toprule
\multicolumn{2}{@{}l}{\textbf{Non-variational}}\\
\cmidrule(lr){1-2}
\textbf{Application} & \textbf{\#qubits} \\
SupermarQ GHZ & 4, 8, 12, 16, 20, 24, 28, 30, 32 \\
SupermarQ HAM & 4, 8, 12, 16, 20, 24, 28, 30, 32 \\
TFIM          & 4, 8, 12, 16, 20, 24, 28, 30, 32 \\
HHL           & 5, 7, 9, 11, 13, 15, 17 \\
\\[-2pt]
\multicolumn{2}{@{}l}{\textbf{Variational}}\\
\cmidrule(lr){1-2}
\textbf{Application} & \textbf{QUBO size} \\
QAOA  & 4, 8, 10, 20, 30 \\
DQAOA & 30 with {\footnotesize \emph{(subqsize\textsuperscript{*}, nsubq\textsuperscript{\dag})}:
         (16,2), (8,4), (12,3)} \\
         & 40 with {\footnotesize \emph{(subqsize\textsuperscript{*}, nsubq\textsuperscript{\dag})}:
         (16,4), (12,4)} \\
\bottomrule
\end{tabular}
{\footnotesize 
    \textsuperscript{*}\,subqsize refers to the sub-QUBO size.
    \quad
    \textsuperscript{\dag}\,nsubq refers to the number of sub-QUBOs.
}
\end{table}

\begin{figure*}[!t]
  \centering

  \makebox[\textwidth][r]{%
    \includegraphics[width=0.96\textwidth]{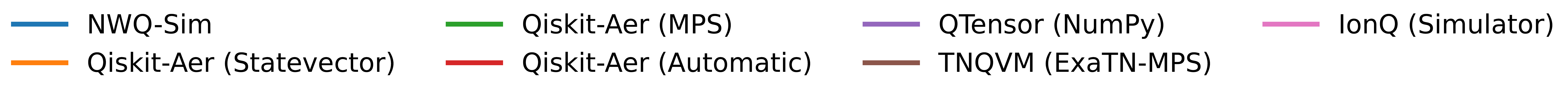}%
  }


  \begin{subfigure}{0.49\textwidth}
    \centering
    \includegraphics[width=\linewidth]{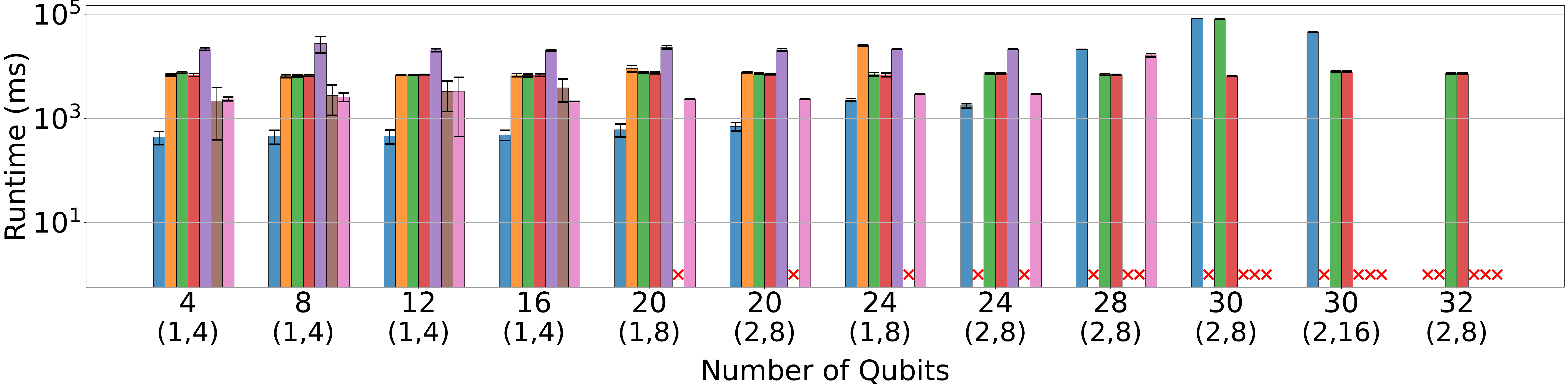}
    \caption{GHZ}
    \label{fig:ghz}
  \end{subfigure}\hfill
  \begin{subfigure}{0.49\textwidth}
    \centering
    \includegraphics[width=\linewidth]{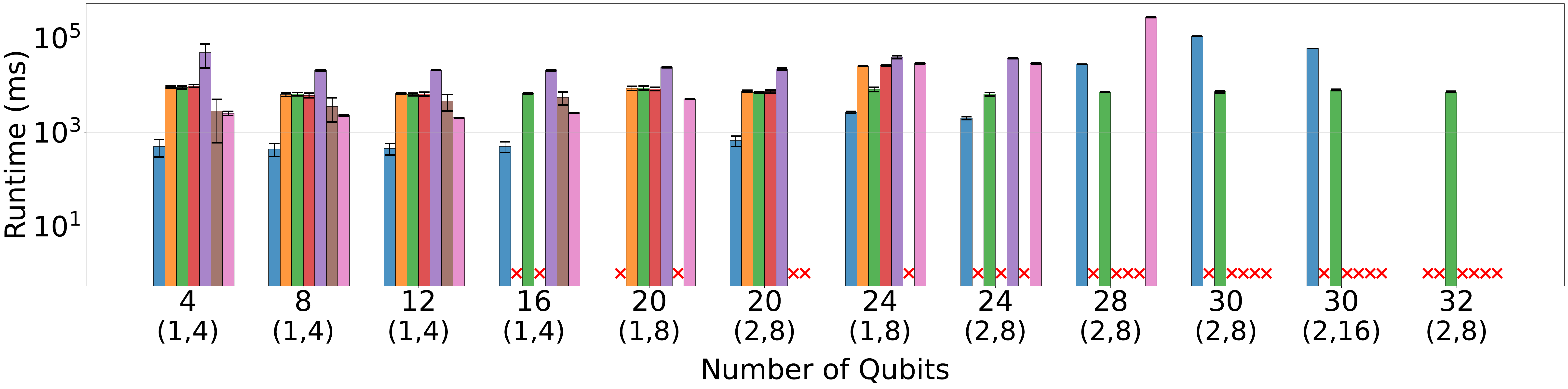}
    \caption{HAM}
    \label{fig:ham}
  \end{subfigure}


  \begin{subfigure}{0.49\textwidth}
    \centering
    \includegraphics[width=\linewidth]{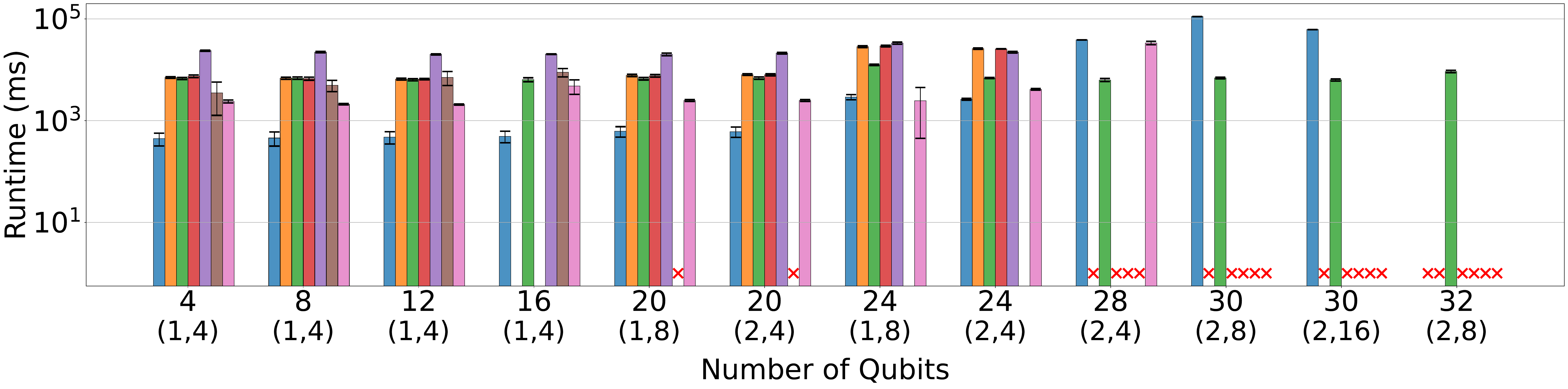}
    \caption{TFIM}
    \label{fig:tfim_scale}
  \end{subfigure}\hfill
  \begin{subfigure}{0.49\textwidth}
    \centering
    \includegraphics[width=\linewidth]{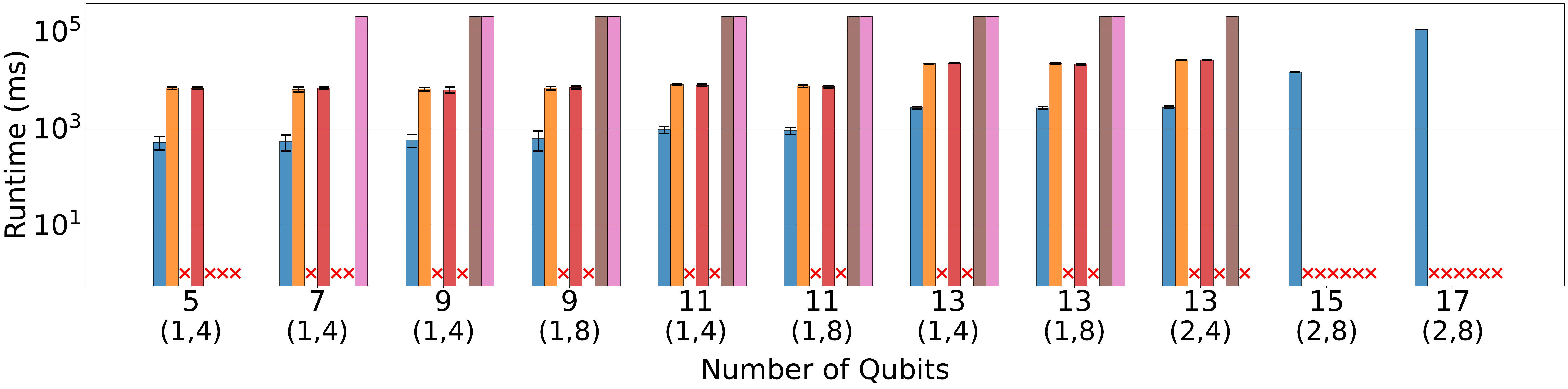}
    \caption{HHL}
    \label{fig:hhl}
  \end{subfigure}


  \begin{subfigure}{0.49\textwidth}
    \centering
    \includegraphics[width=\linewidth]{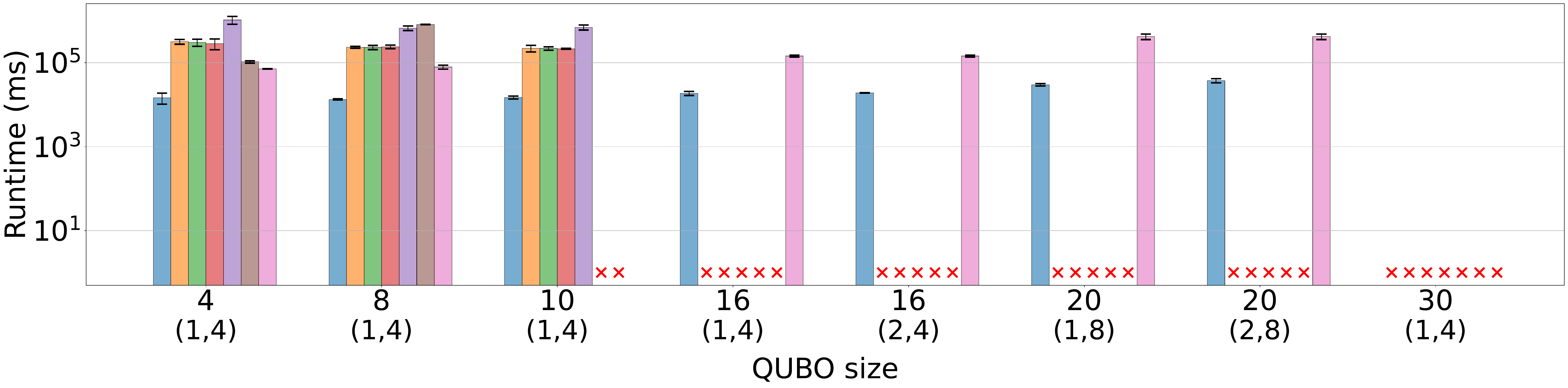}
    \caption{QAOA}
    \label{fig:qaoa}
  \end{subfigure}\hfill
  \begin{subfigure}{0.49\textwidth}
    \centering
    \includegraphics[width=\linewidth]{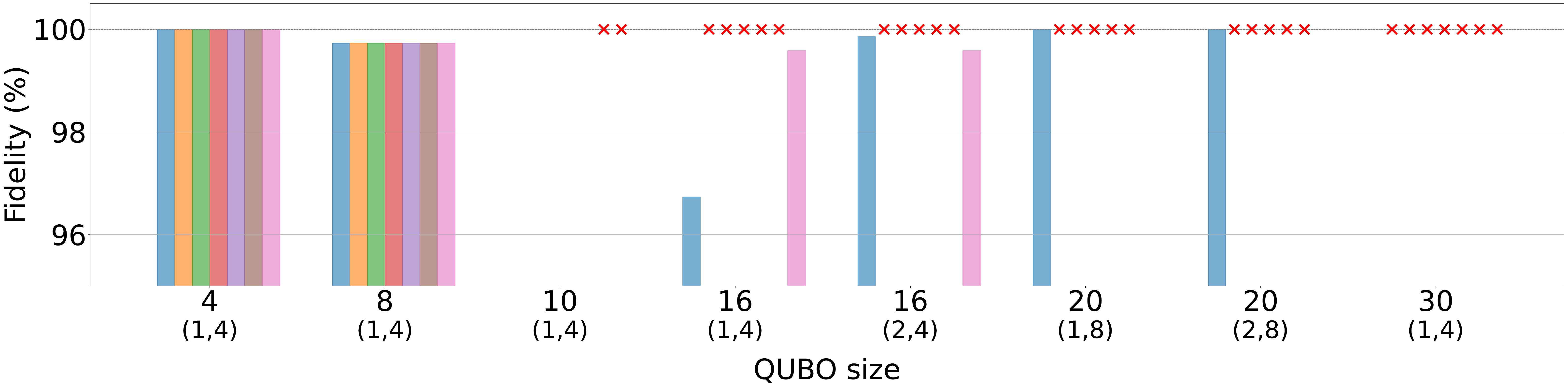}
    \caption{QAOA solution fidelity}
    \label{fig:qaoa_fidelity}
  \end{subfigure}

  \caption{(a)-(e) Results per benchmark across backends. (f) QAOA
    solution fidelity (referenced to D\mbox{-}Wave's hybrid quantum
    annealing solver~\cite{kim2025quantum}).
    This figure represents an approximate weak scaling study where we
    increase both the number of qubits and the number of processes but not exactly proportionally.
    Note: The
      secondary x-axis in each panel denotes \((\#N,\#P)\), i.e.,
      \((\text{num nodes}, \text{processes per node})\).
  }
  \label{fig:benchmarks_all}
\end{figure*}

\section{Results}
\label{sec:results}

We benchmark the integrated backends from Table~\ref{tab:simulators}
on the workloads and sizes listed in Table~\ref{tab:benchmarks}.  Each
point is the mean of three runs with error bars indicating standard
deviation.

\paragraph{Non-variational workloads:}
Figures~\ref{fig:ghz}--\ref{fig:hhl} present runtime scaling for GHZ,
HAM, TFIM, and HHL circuits, respectively.  In GHZ
(Fig.~\ref{fig:ghz}) and HAM from SupermarQ (Fig.~\ref{fig:ham}), all
backends scale to 32 qubits, but runtime separation increases with
size: NWQ-Sim and Qiskit Aer (MPS) remain competitive, while QTensor
slows notably beyond 24 qubits.

For TFIM (Fig.~\ref{fig:tfim_scale}), Qiskit Aer's {\it mps} solver sustains
low runtimes up to 33 qubits, outperforming NWQ-Sim at large sizes.
The TFIM-28 workload, when executed with varying process counts
(an approximate strong scaling study, since the increase
was not strictly proportional).
We observe that state-vector-based simulators such as
NWQ-Sim and Qiskit Aer
exhibit improved performance with
increased resources, whereas MPS-based approaches
do not scale as effectively.

For HHL (Fig.~\ref{fig:hhl}), increasing circuit depth reduces
scalability. NWQ-Sim outperforms at smaller problem sizes, Qiskit Aer
achieves comparable performance for medium instances, and NWQ-Sim
again leads for larger instances, subject to the resource constraints
indicated on the secondary x-axis.

\paragraph{Variational workloads:}
Fig.~\ref{fig:qaoa} and Fig.~\ref{fig:qaoa_fidelity} show QAOA runtime
and fidelity trends as a function of QUBO size.  Runtimes increase
with problem size, with sharper growth when scaling process counts
(\#P) beyond a single LLC domain due to MPI communication overheads.
Missing points (red X) correspond to runs exceeding the two-hour
cutoff.  Fidelity remains consistently above 95\% across tested sizes,
with minor variation due to backend-specific numerical differences.
A red X in the plots indicates a configuration omitted due to
exceeding walltime or resource constraints on the Frontier test
cluster, a scaled-down environment with limited processes per
node. These constraints particularly impact backends such as Qiskit
Aer, which, when run via \texttt{mpi4py}, does not benefit from usual
multi-core optimizations due to not being natively designed for strong
scaling beyond a single node.

\begin{figure}[!htbp]
  \centering
  \includegraphics[width=0.50\linewidth]{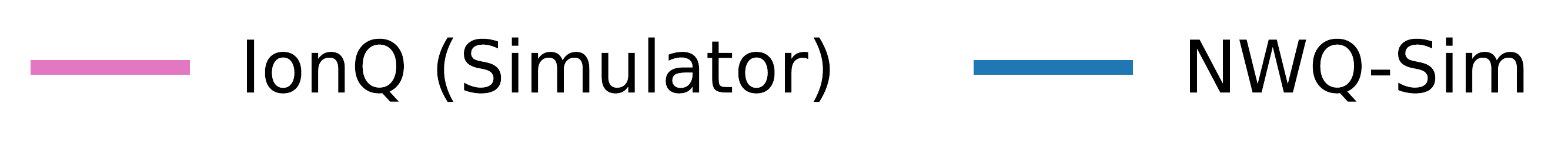}
  \includegraphics[width=0.98\linewidth]{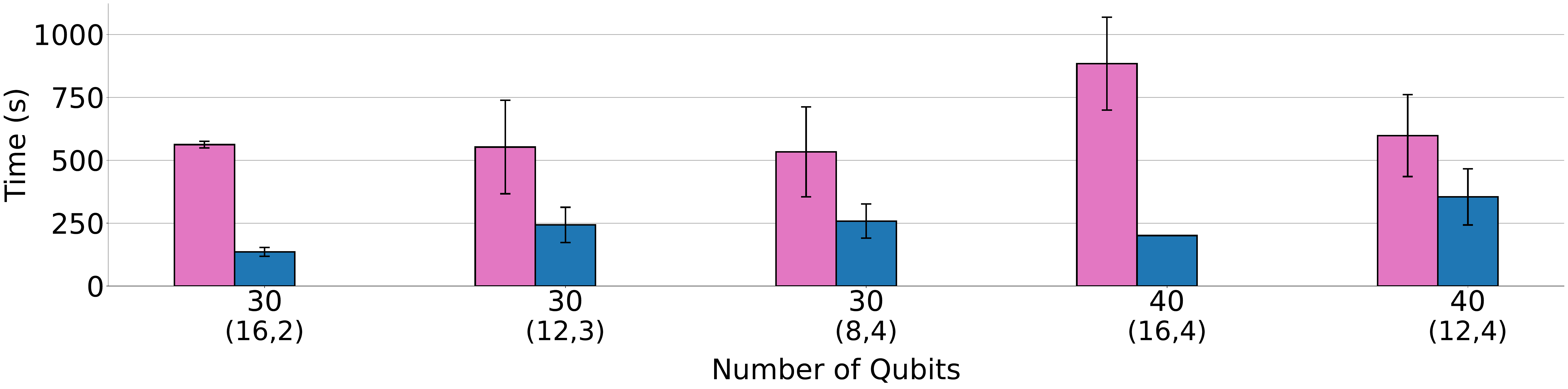}
  \caption{Solving DQAOA using QFw.~\textit{Note: The secondary x-axis
      refers to (subqsize, nsubq) as defined in
      Table~\ref{tab:benchmarks}.}}
  \label{fig:qfw_dqaoa}
\end{figure}

\begin{figure}[!htbp]
    \centering
    \includegraphics[width=0.98\linewidth]{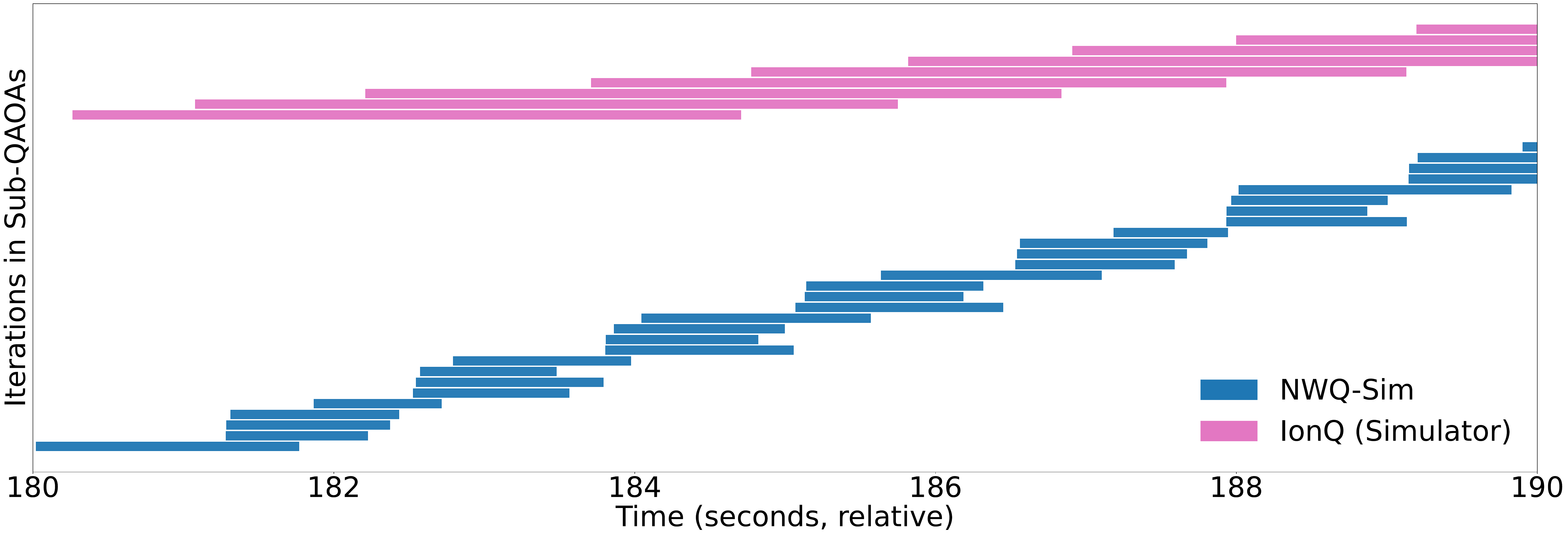}
    \caption{A zoomed portion of the DQAOA-40 run with subqsize=12 and
      nsubq=4.}
    \label{fig:dqaoa_zoomed}
\end{figure}

DQAOA results with NWQ-Sim and IonQ backends are shown in
Figs.~\ref{fig:qfw_dqaoa} and \ref{fig:dqaoa_zoomed}.  The former plot
compares total execution times across different problem sizes and
subproblem configurations, while the latter zoomed view highlights
iteration-level timing patterns for a specific configuration
(subqsize=12, nsubq=4). The zoomed plot clearly shows NWQ-Sim
completing iterations faster (and about four concurrently) and with
more uniform timing compared to the IonQ simulator which involves
calls over the internet and cloud queues.

\section{Discussion}
\label{sec:discussion}

Our evaluation highlights key factors influencing simulator
performance, scalability, and QFw's role in enabling portable quantum
workflows.

No single backend dominates across all problems.  Qiskit Aer's {\it mps}
solver is highly efficient for structured, low-entanglement circuits
such as TFIM, sustaining low runtimes even beyond 30 qubits.
NWQ-Sim's state-vector kernels and MPI parallelization excel for GHZ
and HAM workloads, while QTensor's TTN decomposition favors shallow or
tree-like circuits but slows sharply on deeper or densely connected
topologies.

DQAOA results demonstrate QFw's ability to execute many subQUBOs
concurrently, overlapping RPC communication with backend computation.
Moderate-sized subQUBOs achieve good scaling, but very small subQUBOs
incur fixed overheads from RPC and scheduling, reducing efficiency.
The same application code runs without modification across HPC
simulators and cloud backends, confirming the effectiveness of QFw's
asynchronous, backend-agnostic execution model.

System-level optimizations help reduce OS noise and improve
reproducibility at scale. This includes LLC core reservation, SLURM
heterogeneous job groups, and PRTE-based MPI dispatching.  By fully
decoupling application logic from backend orchestration, QFw enables
fair, repeatable performance comparisons and rapid backend
substitution.

The ability to seamlessly move between local MPI-based simulators and
remote quantum hardware positions QFw as a practical platform for
hybrid execution, large-scale algorithm tuning, and reproducibility
studies in production HPC environments.  This portability is critical
as quantum computing moves toward integrated HPC-quantum deployments.

\section{Conclusion}
\label{sec:conclusion}

We integrated a diverse set of quantum circuit backends into QFw,
making it a hybrid quantum-HPC execution framework that spans multiple
local simulators (NWQ-Sim, Qiskit Aer, TN-QVM, QTensor) via PRTE/MPI
and a remote IonQ service. This unified integration supports both
non-variational and variational workloads, implemented in Qiskit and
PennyLane, and deployable without code changes across heterogeneous
HPC and cloud resources.

Our evaluation on GHZ, HAM, TFIM, and HHL circuits, as well as QAOA,
and DQAOA, confirms that backend performance is highly
workload-dependent: Tensor-network methods such as Qiskit Aer's {\it mps}
excel for structured, low-entanglement problems, while state-vector
engines like NWQ-Sim perform strongly for shallow but highly entangled
workloads. For deeper circuits and large-scale variational algorithms,
distributed MPI execution is essential for scaling, and asynchronous
orchestration enables effective overlap of computation and
communication.

By decoupling application logic from backend orchestration, QFw
enables reproducible, cross-platform benchmarking and rapid backend
substitution, which are critical capabilities for advancing scalable,
portable hybrid quantum-HPC workflows. Future extensions will target real-hardware experimentation, GPU-accelerated tensor-network
backends, automated workload-driven
backend selection, and larger-scale hybrid HPC-cloud studies spanning
multiple hardware targets, further informing backend design and
deployment strategies.

\section*{AI Assistance}
Portions of the text in this manuscript
were refined with the assistance
of the AI tool ChatGPT (OpenAI)~\cite{chatgpt} to
improve clarity and readability.

\begin{acks}
\label{sec:acknowledgment}

This work was supported in part by NSF awards MPS-2531350,
MPS-2410675, PHY-2325080, CISE-2316201, MPS-2120757, CCF-2217020, and
PHY-1818914 as well as DOE DE-SC0025384. This research used resources
of the Oak Ridge Leadership Computing Facility at the Oak Ridge
National Laboratory, which is supported by the Office of Science of
the U.S. Department of Energy under Contract
No. DE-AC05-00OR22725. This material is based upon work supported by
the U.S. Department of Energy, Office of Science, National Quantum
Information Science Research Centers, Quantum Science Center. Notice: This manuscript has been co-authored by UT-Battelle LLC under contract DE-AC05-00OR22725 with the US Department of Energy (DOE). The US government retains and the publisher, by accepting the article for publication, acknowledges that the US government retains a nonexclusive, paid-up, irrevocable, worldwide license to publish or reproduce the published form of this manuscript, or allow others to do so, for US government purposes. DOE will provide public access to these results of federally sponsored research in accordance with the DOE Public Access Plan (\href{http://energy.gov/downloads/doe-public-access-plan}{http://energy.gov/downloads/doe-public-access-plan}).

\end{acks}

\bibliographystyle{unsrtnat}
\bibliographystyle{ACM-Reference-Format}
\bibliography{references}

\end{document}